\documentstyle[prl,aps,graphicx,multicol]{revtex}
  \renewcommand{\narrowtext}{\begin{multicols}{2} \global\columnwidth20.5pc}
  \renewcommand{\widetext}{\end{multicols} \global\columnwidth42.5pc}

\begin{document}
%\twocolumn[\hsize\textwidth\columnwidth\hsize\csname@twocolumnfalse\endcsname

\title{Coherent transfer of Cooper pairs by a movable grain}
\author{L. Y. Gorelik$^{1,2}$, A. Isacsson$^{1,2}$, Y. M. Galperin$^{2,3}$,
 R. I. Shekhter$^{1,2}$ and M. Jonson$^{1,2}$}
\address{$^1$Department of Applied Physics, Chalmers University of
Technology, 412 96 G\"oteborg, Sweden,\\
$^2$Centre for Advanced Study, Drammensveien 78, 0271 Oslo, Norway,\\
$^3$Department of Physics, University of Oslo, P. O. Box
1048, N-0316 Oslo, Norway and
Division of Solid State Physics, Ioffe
Institute of the Russian Academy of Sciences,
St. Petersburg 194021, Russia}
\date{\today}
\maketitle

\begin{abstract}
A coherent hybrid of states with different number of Cooper pairs 
can be built in a superconductor grain as a result of periodically 
repeated discrete encounters with bulk superconductor leads. As a 
direct manifestation of such states a non-dissipative current 
depending on the phase difference between the leads can be measured.  
\end{abstract}
\vspace*{0.1in}
\pacs{PACS number: ????}

\narrowtext

\section{Introduction}
Quantum devices allowing coherent states are in focus of intense
experimental and theoretical studies, in particular in connection
with the problem of quantum computation. A promising example of such
systems is a set of superconducting circuits connected by 
nanometer size contacts.
Phase coherence between two superconductors is maintained if electron
 (Cooper pair)
exchange between them is allowed. A quantum state resulting from 
such coupling is a coherent superposition of 
states with different number of Cooper pairs in the linked 
superconductors. 
As is well known, transfer of Cooper pairs 
is a coherent process causing a ground-state non-dissipative 
current (this is just the Josephson effect\cite{Josephson}).

In recent experiments Nakamura {\it et al.}~\cite{nakamura}
demonstrated that a quantum hybrid of two
charge states can be created by
Josephson tunneling of Cooper pairs between a bulk lead and a 
nanometer-sized superconducting grain, 
commonly referred to as a single-Cooper-pair-box.  
One of the main results of
Ref.~\cite{nakamura} is that this hybrid state remains coherent for a rather
long time, $\ge 20$ ns.  

Nanometer-size
structures  in which charge transfer is coupled to mechanical 
motion have also recently received considerable 
attention. A beautiful
example of charge transport through a C$_{60}$ molecule accompanied by its 
quantum mechanical center-of-mass 
motion has been demonstrated experimentally~\cite{nature_let1} while
another implementation using a macroscopic pendulum to mediate current 
between two spatially separated electrodes was 
provided by Tuominen {\it et al.}~\cite{Touminen}. 
In the latter system the
center-of-mass motion serves as a {\em shuttle} which carries
electrons, as predicted in
Ref.~\cite{chalmers_shuttle}. 

The problem we address here is whether coherent charge transport
is possible by using a movable single-Cooper-pair-box 
oscillating between the two leads thereby providing Cooper pair exchange.
The Josephson current associated with this new mechanism for Josephson
coupling is then presented.

\section{Model system}
First, to demonstrate this phenomenon let us consider a
superconducting grain initially in a state $|n\rangle$ with a fixed number 
of Cooper pairs 
$n$ which  
is brought into contact with a bulk lead and then removed from the
latter. Since the  
Josephson coupling decays strongly with the distance, the grain
finally becomes  
isolated in a state,
$\left|\psi\right>=\sum_n c_n \, e^{-i \phi_n}|n\rangle$. 
The phases $\phi_n$ will then depend on the superconducting phase of
the bulk superconductor.  
Below we will refer to such a state as a {\it Josephson Hybrid} (JH). 
If a grain initially in a hybrid state passes by the lead it acquires a 
charge 
which depends on the coefficients $ c_n \, e^{-i \phi_n}$, most
importantly on the  
phases $\phi_n$, as well as on the phase on the lead. 

Now let us consider a set of leads and allow the grain to pass the
leads sequentially. 
Since between the leads the grain is decoupled
from the environment, the coefficients $c_n$ will remain constant while the 
phases $\phi_n$ 
evolve according to $\dot{\phi}=\hbar^{-1}\langle n|\hat{H}_c|n \rangle$ 
where $\hat{H}_c$ is the Hamiltonian of an isolated grain.
Each interaction with a lead causes a phase-dependent  charge
transfer. The resulting state as well as the total charge transfer is
determined  by the whole ``history'' of the grain's motion.   

An important issue is that the JH can be created only if the
energy differences between states with different numbers of Cooper pairs 
are smaller than the Josephson energy $E_J$.
This can be achieved even if $E_J$ tends to zero by gate electrodes 
which induce charges on the grain. If the gate voltage is properly 
chosen then the states with two subsequent numbers of Cooper pairs on 
the grain become equal. In
other words, the Coulomb blockade of tunneling is lifted, as in the
case of a single-electron transistor. As a result, the JH
consists of only two states with subsequent numbers of Cooper pairs.

As the grain is removed from the lead the problem of how long the ``memory'' 
of a proximity between two
superconductors persists arises. This question was first addressed by 
Leggett in
Ref.~\cite{Leggett} who considered 
the  motion of a superconducting grain between superconducting leads. There it
was assumed that the grain motion in the contact regions is slow
enough to reach a local thermal equilibrium between the grain and the
nearest lead.  
As sources of destruction of such a memory 
he discussed both external random fields and the so-called internal
mutual dephasing due to relaxation to the equilibrium during
subsequent contacts to the leads. 
In contrast to that approach, our system is intrinsically
\emph{non-adiabatic} in the contact regions. Therefore the equilibrium
is not reached during 
the contacts, the system evolves according to the laws of quantum
mechanics, and intrinsic dynamics does not occur.  

Another important difference is that we consider the case of
\emph{strong Coulomb blockade}, when the charging energy $E_C$ is much
larger than the Josephson coupling energy $E_J$. As a
result, the hybrid state is constituted of only two charge states
differing by one Cooper pair as mentioned above. In this 
way we avoid the decoherence due
to quantum beats between many quantum states discussed in
Ref.~\cite{Leggett}.    

\section{Calculations}
\begin{figure}
\centerline{
\includegraphics[width=6cm,angle=-90]{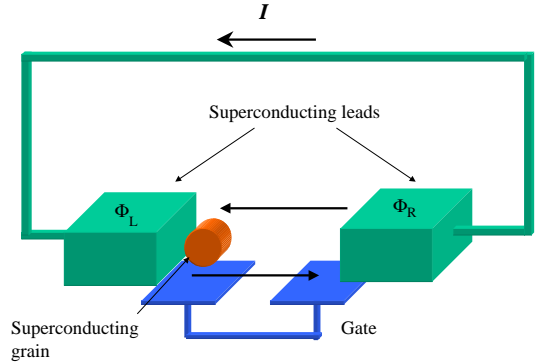}
}
\caption{Schematic of the specific system described in 
the text. A superconducting grain executes periodic motion between two
superconducting bulk leads. The presence of the gate ensures that the
Coulomb blockade is lifted during the contacts between the grain and
superconductor which allows for the grain to be in the Josephson
hybrid state. As the grain moves between the leads Cooper pairs are
shuttled between them creating a DC-current through the structure.}\label{f_fig1}
\end{figure}

For concrete calculations
let us consider charge transfer between two
bulk leads due to repeated alternate contacts of the grain with
those leads  when the grain moves periodically in time as shown in
Fig.~\ref{f_fig1}. A similar system was recently realized experimentally\cite{Blick}.
If we prepare
the system in a state such that after one period its state differs
from the previous one only by a phase factor,   then the charge
transfer is periodic in time. Its magnitude as well as the direction
are determined by the phase difference between the leads, as well as
by the phase   acquired in course of free motion. This is a new
mechanism for Josephson charge transfer due to Cooper pair shuttling
which is a direct manifestation of coherent transport.

Our model Hamiltonian as a function of the grain position $x$ can be 
expressed as a sum of the electrostatic part
\begin{equation}
\hat{H}_C =[e^2/2C(x)]\left[2 {\hat n}+ Q (x)/e \right]
\label{st}
\end{equation}
 and Josephson coupling with the left ($L$) and right ($R$) leads,   
\begin{equation}
\hat{H}_J=-\sum\limits_{s=L,R}%\sum\limits_n
E_J^s(x)\cos(\Phi_s -{\hat \Phi)} \, . \label{j1}
%\left(e^{-i\Phi_s}\left|n+1\right>\left<n\right|+\text{h.c.}\right)\, .
\end{equation}
Here ${\hat n}$ is the operator of the Cooper pair number on the grain, 
$Q(x)$ is the charge 
induced on the grain by the gate, $\Phi_s$ are the order parameter phase of 
the $s$:th lead while 
$\hat \Phi$ is the order parameter phase operator of the grain. The 
operators $\hat n$ and 
$\hat \Phi$ obey the commutation relation $[{\hat n},
{\hat \Phi} ]=i$ as far as we restrict ourselves by the ground state. 
To make that possible we require the typical frequency of the grain 
oscillations to be much less that $\Delta/\hbar$. 
In the Eqs.~(\ref{st}) and (\ref{j1}), 
$C(x)$ is the mutual capacitance, while $E_J^s(x)$ is the Josephson coupling 
energy. Only the states with even number of electrons on the grain are taken 
into account. As it is known \cite{GSJG}, that requires the inequality 
$\Delta \ge e^2/C$. 

The energy of the Josephson coupling between the grain and the electrodes
depends strongly on position and is given by
$E_J^{L,R}(x)=E_0\exp(-\left|\delta x_{L,R}\right|/l)$ 
where $\delta x_{L,R}$ is the distance between the grain and the
right/left contact 
and $l$ is the characteristic tunneling length. 

The dynamics of the system are governed by the %time 
evolution operator
\begin{equation}
\hat{\cal U}(t_2;t_1)=\hat{T}\exp (- i
\hbar^{-1}\int\limits_{t_1}^{t_2}dt^\prime
\hat{H}[x(t^\prime)]).\end{equation}

\begin{figure}[h]
\centerline{
\includegraphics[width=6cm]{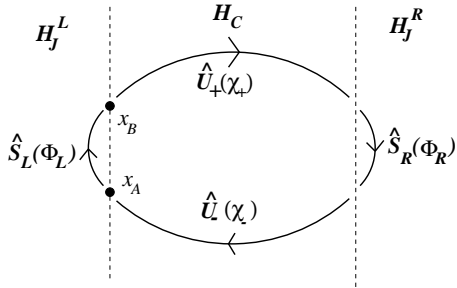}
}
\caption{\label{f_fig3}Schematical trajectory for the system. As the grain
passes the leads the Hamiltonian is completely dominated by the respective 
Josephson term $\hat{H}_J^{L,R}$ while away from the leads the 
charging energy part $\hat{H}_C$ dominates.
This implies that the time evolution operator can be
factored into parts as in Eq. (\ref{ev.}).}
\end{figure}

For periodic grain motion $x(t+T)=x(t)$ it is convenient to consider
the periodic eigenstates 
$\left|\alpha\right>$ at $x_A=x(t_A)$ (cf. with  Fig.~\ref{f_fig3})
\begin{equation}
\hat{\cal U}(t_A+T;t_A)\left|\alpha\right>=
e^{-i\lambda_\alpha}\left|\alpha\right> \, . \label{es}
\end{equation}
The Liouville equation for the statistical operator $\hat{\rho}(t)$ is 
chosen in the form
\begin{equation}
{\rm d}\hat{\rho}(t)/{\rm
d}t =-i\hbar^{-1}[\hat{H}(t),\hat{\rho}(t)] 
-\nu(t)\left[\hat{\rho}(t)-\hat{\rho}_0(t)\right] \, . \label{do} 
\end{equation}
The last item in the right-hand side of Eq.~(\ref{do}) allows for the
relaxation to the adiabatic equilibrium distribution,
$$\hat{\rho}_0=Z^{-1}\exp\left(- \beta \hat{H} [x(t)]\right).$$ 
Here $\beta$ is the inverse temperature
while $Z={\rm Tr}\, \exp\left(- \beta \hat{H} [x(t)]\right)$.
We specify the relaxation as 
$%\begin{equation}
\nu(t)=\nu_0\exp\left[-\left|\delta x (t)\right|/l\right]
$. %\, . \end{equation}
One needs to introduce the relaxation in order to ``forget'' the
initial conditions to Eq.~(\ref{do}). Physically, the relaxation is
caused by incoherent charge transfer, such as quasiparticle
tunneling. In the following we will assume 
that the relaxation is infinitely slow. Then it does not enter the
results explicitly. However, it determines the populations
through the balance equation (\ref{do}). 
The latter has the simplest form in the
representation of the evolution operator, Eq.~(\ref{es}). In this
case $\hat{\rho}$ is diagonal in the $|\alpha\rangle$ representation  
\begin{equation}
{\rho}_{\alpha \alpha}=\overline{\nu(t)\left<\alpha\right|\hat{\cal U}(T ;t)
\hat{\rho}(t)\hat{\cal U}^\dagger(T;t)\left|\alpha\right>}/
{\overline{\nu}} \label{av}
\end{equation}
where $\overline{A(t)}=T^{-1}\int\limits_0^T dt A(t)$ is the time
average over the period $T$ of the grain's motion.
Equation (\ref{av}) can be directly obtained as a time average of  
Eq.~(\ref{do}).   
Knowing the eigenstates (\ref{es}) one can easily calculate the
electric current 
\begin{equation}
{\bar I}=2ef \sum_\alpha \rho_{\alpha\alpha }I_\alpha\, , \quad f
\equiv T^{-1}
\label{current}
\end{equation}
where $I_\alpha$ is the charge carried through the cross section
$x_A$ (cf. with  Fig.~\ref{f_fig3}) by the state $|\alpha\rangle$ during 
one cycle and
is given by the general relation
\begin{equation}
I_\alpha\equiv\left<\alpha\right|\hat{n}
-\hat{\cal U}^\dagger(t_B;t_A)\hat{n}\hat{\cal U}(t_B;t_A)\left|\alpha\right>
={\partial \lambda_\alpha}/{\partial \Phi_L}\, .
\end{equation}
The dynamics of the system under consideration is most simple if the
electrostatic energy cost per tunneling  Cooper pair, $E_C \equiv
(2e)^2/2C$, is larger than the Josephson coupling energy, $E_J$. Then
the tunneling transport can be blocked by Coulomb effects, and
resonant tunneling of Cooper pairs is possible only at very specific
values of the induced charge $Q$. In particular, when 
$Q=Q_n=-(2n+1)e$ only two states with the number of excess Cooper
pairs at the grain which differs by one are possible. This leaves us
inside a two-state Hilbert space which can be characterized by the
Cooper pair number at the grain, $|n\rangle$ and $|n+1\rangle$. If at
that instant the Josephson coupling is ``on'', a coherent hybrid,
$c_1|n\rangle +c_2e^{i\phi}|n+1\rangle$, is formed.  Such a situation
takes place each time the grain passes the lead. Consequently, the
state evolution can be decoupled into the ``scattering'' regions where
the grain is in touch with a lead and ``free motion'' regions where
the grain is decoupled from the leads and its state evolves as that of
a non-interacting system. The grain's trajectory is schematically
shown in Fig.~\ref{f_fig3}. For such motion the time evolution
operator $\hat{\cal U}$
%(\ref{evolution})  
has the form of a $2 \times 2$ matrix
which can be factored into
four parts pertaining to the two scattering events and the evolution
 of the state between the leads i.e.
\begin{equation}
\hat{\cal U}(t_A+T;t_A)=\hat{U}_-\hat{S}_R\hat{U}_+\hat{S}_L.
\label{ev.}
\end{equation}
Here 
\begin{equation}
\hat{U}_\pm=\exp(i\chi_\pm\sigma_3)\, \mbox{ , } \chi_\pm =(1/\hbar)\int dt
\, \delta E_c[x(t)]
\end{equation}
describe the ``free'' evolution along the upper, right moving,  and
the lower, left moving, paths respectively.
The quantity $\delta E_c(x)$ is the
difference in electrostatic energy for the two charge states in the
hybrid while $\sigma_3$ is the Pauli matrix.
The scatterings with the leads are given by the matrices 
\begin{equation}
\hat{S}_{s}=\exp \left[ i
\vartheta_{s}\, (\sigma_1\cos \Phi_s+\sigma_2\sin \Phi_s) \right]
.\end{equation} 
Here $\vartheta_{s}$ %is determined by by 
are dimensionless {\em contact times}  
with the leads $s=R,L$ defined as
$%\begin{equation}
\vartheta_{s}=\hbar^{-1}\int dt\, E_J^s(t)$. 
In the following we will assume for simplicity that
$\vartheta_R=\vartheta_S=\vartheta$.  
Since the the exponential rapidly decays when the grain is far from
the lead the integral can be decoupled into the integrals over the
scattering regions and the integrals over the
regions of the free motion. 
Inserting Eq. (\ref{ev.}) for the evolution operator into the
expression (\ref{current}) for the electric current, one finds the  
relationship between the dc current, $\bar I$ and the eigenvalues of the  
evolution operator, 
\begin{equation}
\overline{I}=2ef\sum_i \rho_{ii}I_i = ef\left(
{\partial \lambda_1}/{\partial \Phi_R}\right)
\left({\partial \lambda_1}/{\partial \vartheta}\right)\, .
\end{equation}
This expression is non-trivial because it expresses the current in a
non-equilibrium state in an equilibrium fashion.

The eigenvalues  for our two-state
system can be expressed as $\exp(-i\lambda_\alpha)$, where
$\lambda_1=-\lambda_2 \equiv \lambda$ is given   
by the equation 
\begin{equation}
\cos\lambda=\cos^2 \vartheta \, \cos\chi-\sin^2\vartheta\, \cos\Phi\,
.\label{est} 
\end{equation} 
Here we have defined $\chi=\chi_-+\chi_+$ and
$\Phi=(\Phi_R-\Phi_L)+(\chi_+-\chi_-)$. 
The dc current for the case $\vartheta_L=\vartheta_R \equiv
\vartheta$ is
\begin{equation}
{\bar I} =2ef\frac{\cos \vartheta \, \sin^3 \vartheta \,
\sin\Phi \, (\cos\chi+\cos\Phi)}{1-(\cos^2 \vartheta\,
\cos\chi-\sin^2\vartheta\, \cos\Phi)^2}\, . \label{cur} 
\end{equation}

\section{Discussion and conclusions}

The oscillating dependence of the dc current  on the
phase difference $\Phi_R - \Phi_L$ is a direct
consequence of the fact that the coherent states are controlled by the
phase difference $\Phi$, see Eq.~(\ref{est}). In this way the Cooper
pair shuttle discussed above provides a way to prepare and control a
coherent superposition of quantum states in  a two-state system which
is referred to as a qubit. If there is no phase difference, $\Phi_L=\Phi_R $,
but the grain's trajectory is {\em asymmetric}, $\chi_+ \ne 
\chi_-$, the current still does not vanish. If the grain's trajectory
embeds some magnetic flux created by external magnetic field with
vector-potential ${\bf A} ({\bf r})$, an extra item
$(2\pi/\Phi_0)\oint {\bf A} ({\bf r})\cdot d{\bf r}$ enters the
expression for the phase difference $\Phi$ which must be gauge-invariant.  
Here $\Phi_0=\pi \hbar c/e$ is the magnetic flux quantum.

\begin{figure}[h]
\centerline{
\includegraphics[width=7cm]{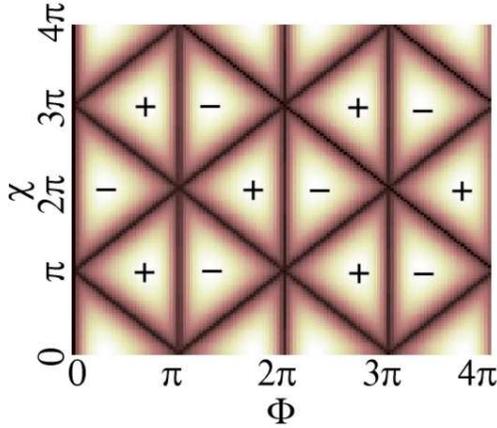}
}
\caption{\label{f_fig4}The magnitude of the current 
$\overline{I}$ in Eq. (\ref{cur}) in units of
$I_0=2ef$ as a function of the phases $\Phi$ and $\chi$. Regions of black
correspond to no current and regions of white to $|\overline{I}/I_0|=0.5$.
The direction of the current is indicated in the figure by signs $\pm$.
To best see the triangular structure of the current the contact time
has been chosen to $\vartheta=\pi/3$. $I_0$ 
contains only the fundamental 
frequency of the grain's motion and the Cooper pair charge.}
\end{figure}
At small contact times, $\vartheta \ll 1$, 
\begin{equation} %\label{}
{\bar I} \approx 2ef \vartheta^3 \sin \Phi \,(\cos \chi + \cos \Phi)
.\end{equation}
The smallness of this current in this case is natural because it is the
contact with the leads which creates the coherent state.

To make the system under consideration realistic, one has to design a
setup with large decoherence times. Firstly, the device should be
carefully screened from electromagnetic perturbations as well as from
other time-dependent magnetic fields. Secondly, the gate electrode, as
well as the insulating region between the gate and the grain, should
contain as little as possible of charged dynamic defects. 
Time-dependent switching between the states of those
defects would produce charge fluctuations coupled to the charge states
of the grain. It seems very difficult to estimate theoretically 
the total dephasing. For the estimate of the dephasing one can take the experimental results
obtained by Nakamura {\em et al.}~\cite{nakamura} who have observed
coherent response of a superconducting qubit during the time $\tau_\phi =
10^{-9}-10^{-8}$ s. This is a lower bound for our case because
the grain is coupled to the leads only during a small part of the
period of its motion. Consequently, the contact time is of the order
of $\tau_c \approx T\sqrt{l/d}$ where $d$ is the distance between the
leads. It is this quantity that should be compared to the dephasing
time, $\tau_\phi$, since the rest of the period the grain is decoupled
from the leads.  One can imagine
another implementation of the Cooper pair shuttling where the alternating
couplings between the grain and the leads are controlled by time dependent
potential barriers rather than by mechanical motion. 
To conclude, we 
have demonstrated a possibility to create a coherent quantum
superposition between two charge states of a superconducting grain by
repeated alternating contacts with two (or more) superconducting
leads. This state can be controlled by the phase difference between
the leads and monitored by a phase-dependent dc current though the
closed circuit including the leads.

\widetext
\end{document}